\documentclass{article}

\usepackage{PRIMEarxiv}

\usepackage[utf8]{inputenc} 
\usepackage[T1]{fontenc}    
\usepackage{hyperref}       
\usepackage{url}            
\usepackage{booktabs}       
\usepackage{amsfonts}       
\usepackage{nicefrac}       
\usepackage{microtype}      
\usepackage{lipsum}
\usepackage{fancyhdr}       
\usepackage{graphicx}       
\graphicspath{{media/}}     

\usepackage{enumitem}
\usepackage{booktabs}
\usepackage{siunitx}
\usepackage[table]{xcolor}
\usepackage{tabularx}
\usepackage{caption}
\usepackage{xcolor}
\usepackage{adjustbox}
\usepackage{parskip}
\usepackage{array}
\usepackage{booktabs}
\usepackage{multirow}
\usepackage{multicol}
\usepackage{tabularx}
\usepackage{graphicx}

\title{TransRUPNet for Improved Polyp Segmentation}

\author{Debesh Jha$^{1}$, Nikhil Kumar Tomar$^{1}$,  Debayan Bhattacharya$^{2}$, Koushik Biswas$^{1}$, Ulas Bagci$^{1}$ \\
Machine \& Hybrid Intelligence Lab, Department of Radiology, Northwestern University\\
Institute of Medical Technology and Intelligent System, Hamburg University of Technology\\ \&  Clinic for Ears, Nose and Throat, University Medical Center Hamburg-Eppendorf, Germany.}

\begin{document}
\maketitle

\begin{abstract}
Colorectal cancer is among the most common cause of cancer worldwide. Removal of precancerous polyps through early detection is essential to prevent them from progressing to colon cancer. We develop an advanced deep learning-based architecture, Transformer based Residual Upsampling Network (TransRUPNet) for automatic and real-time polyp segmentation. The proposed architecture, TransRUPNet, is an encoder-decoder network consisting of three encoder and decoder blocks with additional upsampling blocks at the end of the network. With the image size of $256\times256$, the proposed method achieves an excellent real-time operation speed of \textbf{47.07} frames per second with an average mean dice coefficient score of 0.7786 and mean Intersection over Union of 0.7210 on the out-of-distribution polyp datasets. The results on the publicly available PolypGen dataset suggest that TransRUPNet can give real-time feedback while retaining high accuracy for in-distribution datasets. Furthermore, we demonstrate the generalizability of the proposed method by showing that it significantly improves performance on out-of-distribution dataset compared to the existing methods. The source code of our network is available at \url{https://github.com/DebeshJha/TransRUPNet}.
\end{abstract}


\section{INTRODUCTION}
Colonoscopy is widely considered the gold standard for the diagnosis of colon cancer. Early detection of polyp is important as even a small increase in adenoma detection rate can significantly decrease interval colorectal cancer incidence~\cite{urban2018deep}. Studies report a polyp miss rate of around 22-28\%~\cite{leufkens2012factors}. There are several reasons for polyp miss-rates in colonoscopy, for example, the skill of endoscopists, bowel preparation quality, fast withdrawal time, visibility, and differences in polyp characteristics.

Deep learning-based algorithms have emerged as a promising approach to improve diagnostic performance by highlighting the presence of precancerous tissue in the colon and reducing the clinical burden. OOD detection and generalization are essential for developing computer-aided diagnostic support systems in colonoscopy.  The reliability and safety of deep learning models are important. Traditional deep learning models are trained based on closed-world assumption, where the test dataset is considered from the same distribution as the training data. Therefore, even the well performing model may fail on OOD samples.

We extend our study by training on dataset from one center and testing on dataset from different countries that may have distinct distribution as compared to the data used for training models. 
In this study, we introduce TransRUPNet architecture to address the critical need for clinical integration of a real-time and highly accurate polyp segmentation routine. The main contributions of this work are as follows:

\begin{enumerate}
\item We propose TransRUPNet, an encoder-decoder architecture specifically designed for accurate, real-time and improved polyp segmentation, emphasizing high performance on diverse external datasets. 

\item We compared the performance of TransRUPNet with the existing state-of-the-art (SOTA) methods in four different polyp datasets (one within training distribution and three OOD datasets) to show the method's superiority.

\item Our architecture showed strong generalization capabilities, outperforming 10 SOTA methods in terms of segmentation performance and adaptability.

\end{enumerate}

\begin{figure*}[!t]
    \centering
    \includegraphics[width =0.99\textwidth]{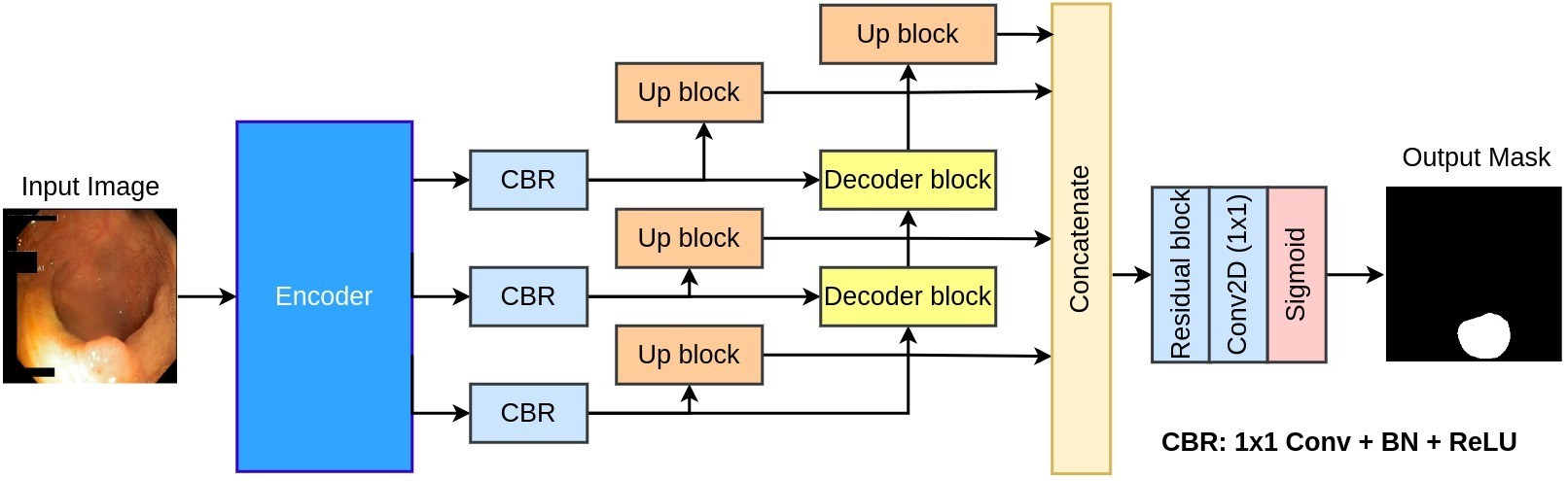}
    \caption{Overall architecture of the TransRUPNet.}
    \label{fig:rupnet}
    \vspace{-5mm}
\end{figure*}

\section{Related Work}
Recently, there has been a significant advancement in the development of models for polyp segmentation. While U-Net based architectures have been widely used, several other approaches have also been proposed that focus on capturing boundary details and leveraging the camouflage property of polyps. One such architecture is PraNet~\cite{fan2020pranet}, which incorporates reverse attention modules to incorporate boundary cues. It combines a global feature map obtained using a parallel partial decoder. Another approach proposed by~\cite{yue2022boundary} introduces a boundary constraint network that utilizes a bilateral boundary extraction module to analyze polyp and non-polyp regions. Polyp-PVT~\cite{dong2021polyp} takes a different approach by introducing a camouflage identification module with a pyramid vision transformer (PVT) encoder. This module aims to capture polyp cues that are concealed in low-level features.

The success of transformer-based approaches in polyp segmentation has led to the development of more similar works in the field. ColonFormer~\cite{duc2022colonformer} proposes a hierarchical transformer combined with a hierarchical pyramid network, incorporating a residual axial attention module for efficient polyp segmentation. Besides polyp segmentation, there are medical image segmentation architectures and techniques that have shown promising performance on radiology images~\cite{jha2024ct,zhou2018unet++}.  Overall, this research demonstrates a wide range of architectural variations and techniques used for polyp segmentation, inspiring further research for the computer-aided diagnosis system for colon polyp segmentation. 


\section{Method}
Figure~\ref{fig:rupnet} shows the block diagram of the proposed TransRUPNet architecture. Our architecture is designed with a primary focus on achieving high-performance metrics and real-time speed, which are essential for routine colonoscopy examinations. Inspired by a recent transformer-based network, PVTFormer~\cite{jha2024ct}, which showed SOTA performance on liver segmentation tasks, we aim to solve the critical challenge in polyp segmentation.

Our architecture follows an encoder decoder scheme that begins with a Pyramid Vision Transformer (PVT)~\cite{wang2021pyramid} as a pre-trained encoder. We leverage a  PVT model (pvt\_v2\_b2) which is pretrained on ImageNet~\cite{5206848} classification task to initialize the encoder weights. We extract three different feature maps from the encoder, which have rich hierarchical features learned by the transformer model, and pass them through a series of $1\times1$ Conv, Batch Normalization, and ReLU activation for reducing the number of feature channels to $64$. The reduced feature maps are then passed to the up block and the decoder blocks. Within the up block, the input feature map is first passed through a bilinear upsampling to upscale the feature map's height and width to that of the original input image. Next, the upsampled feature map is passed through a residual block to learn a more robust representation. 

The decoder block also begins with a bilinear upsampling layer to increase the spatial dimensions by a factor of $2$ and then concatenates with the reduced feature from the encoder. Next, the concatenated feature map is passed through a residual block to learn more robust semantic features that help generate a fine-quality segmentation mask. The output from the first decoder block is passed to the next decoder block, which is further followed by an up block. We concatenate the output from all four up blocks into a single feature representation. After that, the concatenated feature map is followed by a residual block, $1\times1$ convolution and a sigmoid activation to generate the final polyp segmentation mask.

\begin{table*}[t!]
\centering
\caption{Quantitative results on the Kvasir-SEG test dataset.}
 \begin{tabular} {l|c|c|c|c|c|c|c}
\toprule
\textbf{Method}  &\textbf{Backbone} & \textbf{mIoU}  &\textbf{mDSC}  &\textbf{Recall}& \textbf{Precision} &\textbf{F2} &\textbf{FPS}\\ 
\hline

U-Net~\cite{ronneberger2015u} &	-&	0.7472&	0.8264&	0.8504&	0.8703&	0.8353 &\textbf{106.88} \\
U-Net++~\cite{zhou2018unet++}&	-&	0.7420&	0.8228&	0.8437&	0.8607&	0.8295 & \textcolor{red}{81.34}\\
ResU-Net++~\cite{jha2019resunet++}&	-&	0.5341&	0.6453&	0.6964&	0.7080&	0.6576 & 43.11\\
HarDNet-MSEG~\cite{huang2021hardnet}&	HardNet68&	0.7459&	0.8260&	0.8485&	0.8652&	0.8358 &34.80\\
ColonSegNet~\cite{jha2021real}&	-&	0.6980&	0.7920&	0.8193&	0.8432&	0.7999 &73.95 \\
DeepLabV3+~\cite{chen2018encoder} &ResNet50 &0.8172 &0.8837 &0.9014 &0.9028 &0.8904 &67.88 \\
PraNet~\cite{fan2020pranet} &Res2Net &0.8296 &0.8942 &0.9060 &\textcolor{red}{0.9126} &0.8976 &31.89\\
TGANet~\cite{tomar2022tganet}&	ResNet50 & \textcolor{red}{0.8330}&\textcolor{red}{0.8982}& \textcolor{red}{0.9132}&	{0.9123}&	\textcolor{red}{0.9029} &36.58\\
TransResU-Net\cite{tomar2022transresu} &ResNet50 &0.8214 &0.8884 &0.9106 &0.9022 &0.8971 & 48.61 \\
TransNetR~\cite{jha2023transnetR} & ResNet50 & 0.8016 & 0.8706 & 0.8843 & 0.9073 & 0.8744 & 54.60 \\

\textbf{TransRUPNet (Ours)} & PVT &\textbf{0.8445} &\textbf{0.9005}& \textbf{0.9195}& \textbf{0.9170}& \textbf{0.9048} & 47.07\\
\bottomrule
\end{tabular}
\label{tab:resultiD}
\vspace{-3mm}
\end{table*}

\begin{table*}[t!]
\centering
\caption{Quantitative results on the Kvasir-SEG test dataset.}
 \begin{tabular} {l|c|c|c|c|c|c}
\toprule
\textbf{Method}  &\textbf{Backbone} & \textbf{mIoU}  &\textbf{mDSC}  &\textbf{Recall}& \textbf{Precision} &\textbf{F2}\\ 

\hline\multicolumn{7}{@{}l}{\textbf{Training dataset: Kvasir-SEG -- Test dataset: PolypGen (C6)}}
\\  \hline
U-Net~\cite{ronneberger2015u} &	- &0.5384 &0.6126 &0.7054 &0.7508 &0.6362 \\
U-Net++~\cite{zhou2018unet++}&	- &0.5355 &0.6163 &0.7340 &0.7230 &0.6564 \\
ResU-Net++~\cite{jha2019resunet++}&	- &0.2816 &0.3684 &0.6220 &0.3526 &0.4326 \\
HarDNet-MSEG~\cite{huang2021hardnet}&	HardNet68 &0.5548 &0.6341 &0.7197 &0.7722 &0.6487\\
ColonSegNet~\cite{jha2021real}&	- &0.4410 &0.5290 &0.6199 &0.6403 &0.5424\\
DeepLabV3+~\cite{chen2018encoder}&	ResNet50 &\textcolor{red}{0.7031} &\textcolor{red}{0.7629} &\textcolor{red}{0.7773} &0.8693 &\textcolor{red}{0.7674}\\
PraNet~\cite{fan2020pranet}&	Res2Net &0.6691 &0.7307 &0.7612 &{0.8755} &0.7378\\
TGANet&	ResNet50 &0.6750 &0.7382 &0.7692 &{0.8887} &0.7391\\
TransResU-Net\cite{tomar2022transresu} &ResNet50 &0.6907 &0.7466 &0.7443 &\textcolor{red}{0.9086} &0.7434 \\
TransNetR~\cite{jha2023transnetR} & ResNet50 &0.6336 &0.6919 &0.6784 &\textbf{0.9432} &0.6805 \\
\textbf{TransRUPNet (Ours)} & PVT &\textbf{0.7210} &\textbf{0.7786} &\textbf{0.8522} &0.8175 &\textbf{0.7929}\\

\hline\multicolumn{7}{@{}l}{\textbf{Training dataset: Kvasir-SEG -- Test dataset: CVC-ClinicDB}}
\\  \hline
U-Net~\cite{ronneberger2015u}&	-&	0.5433&	0.6336&	0.6982&	0.7891&	0.6563 \\
U-Net++~\cite{zhou2018unet++}&	-&	0.5475&	0.6350&	0.6933&	0.7967&	0.6556 \\
ResU-Net++~\cite{jha2019resunet++}&	-&	0.3585&	0.4642&	0.5880&	0.5770&	0.5084 \\
HarDNet-MSEG~\cite{huang2021hardnet}&	HardNet68&	0.6058&	0.6960&	0.7173&	0.8528&	0.7010 \\
ColonSegNet~\cite{jha2021real}&	-&	0.5090&	0.6126&	0.6564&	0.7521&	0.6246 \\
DeepLabV3+~\cite{chen2018encoder}&	ResNet50&	0.7388&	0.8142&	0.8331&	0.8735&	0.8198 \\
PraNet~\cite{fan2020pranet}&	Res2Net&	0.7286&	0.8046&	0.8188&	\textcolor{red}{0.8968}&	0.8077 \\
TGANet &ResNet50 & \textcolor{red}{0.7444} &\textcolor{red}{0.8196} &0.8290 &{0.8879} &\textcolor{red}{0.8207}\\

TransResU-Net\cite{tomar2022transresu} &ResNet50 &0.7342 &0.8082 &\textcolor{red}{0.8331} &0.8861 &0.8173\\
TransNetR~\cite{jha2023transnetR} & ResNet50 &0.6912 &0.7655 &0.7570 &\textbf{0.9201} &0.7565 \\
\textbf{TransRUPNet (Ours)} & PVT &\textbf{0.7765} &\textbf{0.8539} &\textbf{0.8736} &0.8870 &\textbf{0.8590} \\

\hline\multicolumn{7}{@{}l}{\textbf{Training dataset: Kvasir-SEG -- Test dataset: BKAI-IGH}}
\\  \hline
U-Net~\cite{ronneberger2015u} &	- &0.5686 &0.6347 &0.6986 &0.7882 &0.6591 \\
U-Net++~\cite{zhou2018unet++}&	- &0.5592 &0.6269 &0.6900 &0.7968 &0.6493 \\
ResU-Net++~\cite{jha2019resunet++}&	- &0.3204 &0.4166 &0.6979 &0.3922 &0.5019 \\
HarDNet-MSEG~\cite{huang2021hardnet}&HardNet68 &0.5711 &0.6502 &0.7420 &0.7469 & 0.6830 \\
ColonSegNet~\cite{jha2021real}&	- &0.4910 &0.5765 &0.7191 &0.6644 &0.6225 \\
DeepLabV3+~\cite{chen2018encoder}&	ResNet50 &0.6589 &0.7286 &0.7919 &0.8123 &\textcolor{red}{0.7493} \\
PraNet~\cite{fan2020pranet} &Res2Net &0.6609 &\textcolor{red}{0.7298} &\textcolor{red}{0.8007}	&0.8240 &0.7484 \\
TGANet&	ResNet50 &\textcolor{red}{0.6612} &0.7289 &0.7740 &0.8184 &0.7412\\

TransResU-Net\cite{tomar2022transresu} &ResNet50 &0.6457 &0.7067 &0.7363 &\textcolor{red}{0.8635} &0.7148\\
TransNetR~\cite{jha2023transnetR} & ResNet50 &0.5998 &0.6601 &0.6660 &\textbf{0.9072} &0.6583 \\
\textbf{TransRUPNet (Ours)} & PVT &\textbf{0.7218} &\textbf{0.7945} &\textbf{0.8497} &0.8337 &\textbf{0.8072}\\
\bottomrule
\end{tabular}
\label{tab:resultOOD}
\vspace{-5mm}
\end{table*}

\section{Experiment}
\subsection{Dataset}
We use four publicly available colonoscopy polyp segmentation datasets, namely, Kvasir-SEG~\cite{jha2020kvasir}, PolypGen~\cite{ali2021polypgen}, BKAI-IGH~\cite{lan2021neounet}, and CVC-ClinicDB~\cite{bernal2015wm}. Kvasir-SEG was collected from Norway. PolypGen dataset was collected from 6 medical centers, Norway, Italy, France, the United Kingdom, and Egypt, incorporating more than 300 patients. It is a complex dataset containing diverse samples from different cohort populations from six countries. BKAI-IGH was collected in Vietnam, and CVC-ClinicDB was collected in Spain. We use the Kvasir-SEG for in-distribution testing and PolypGen, BKAI-IGH, and CVC-ClinicDB for OOD generalization testing.  

\subsection{Experiment setup and configuration}
We select Kvasir-SEG~\cite{jha2020kvasir} dataset for training all the models. It contains 1000 images and mask pair. We use 880 images and masks for training our method and the rest for validation and testing. In addition, we perform extensive data augmentation to increase the size of training samples. All the experiments are implemented using with PyTorch framework. We run all the experiments on an NVIDIA RTX 3090 GPU system. We use Adam optimizer with a learning rate of 1e$^{-4}$ and a batch size of 8. Additionally, we use a combined binary cross-entropy and dice loss for training our models. 

\section{Result}
\textbf{Comparison with SOTA on in-distribution data: } Table~\ref{tab:resultiD} shows the result of the TransRUPNet and other benchmarking algorithms used in the study. It obtained a mean dice coefficient of 0.9005, mIoU of 0.8445, recall of 0.9195, precision of 0.9170, and F2-score of 0.9048. With the image resolution of $256\times256$, TransRUPNet obtained a real-time processing speed of 47.07 frames per second (FPS). The most competitive network to TransRUPNet was TGANet, to whom our architecture outperformed by 1.15\% in mIoU and 0.23\% in DSC. The processing speed of our network is almost 1.5 times that of TGANet.

\textbf{Comparison with SOTA on OOD data: } We have evaluated the performance of  TransRUPNet on three OOD datasets. We have highlighted the best and second-best scores in Table~\ref{tab:resultOOD}. For this, we train different models on Kvasir-SEG dataset and test it on PolypGen (Center 6). Kindly note that this is the experimental setup for EndoCV 2021 Challenge~\cite{ali2022assessing}. We obtained an improvement of 4.6\% in mIoU and 4.04\% in mDSC as compared to TGANet. Similarly, we obtained an improvement of 3.21\% in mIoU and 3.43\% in mDSC on CVC-ClinicDB datasets. Additionally, we obtained an improvement of 6.06\% in mIoU and 6.56\% in mDSC for the TransRUPNet when tested on BKAI-IGH datasets as compared to the SOTA TGANet~\cite{tomar2022tganet}. 

\begin{figure}[!t]
\centering
\includegraphics[width=0.9\linewidth]{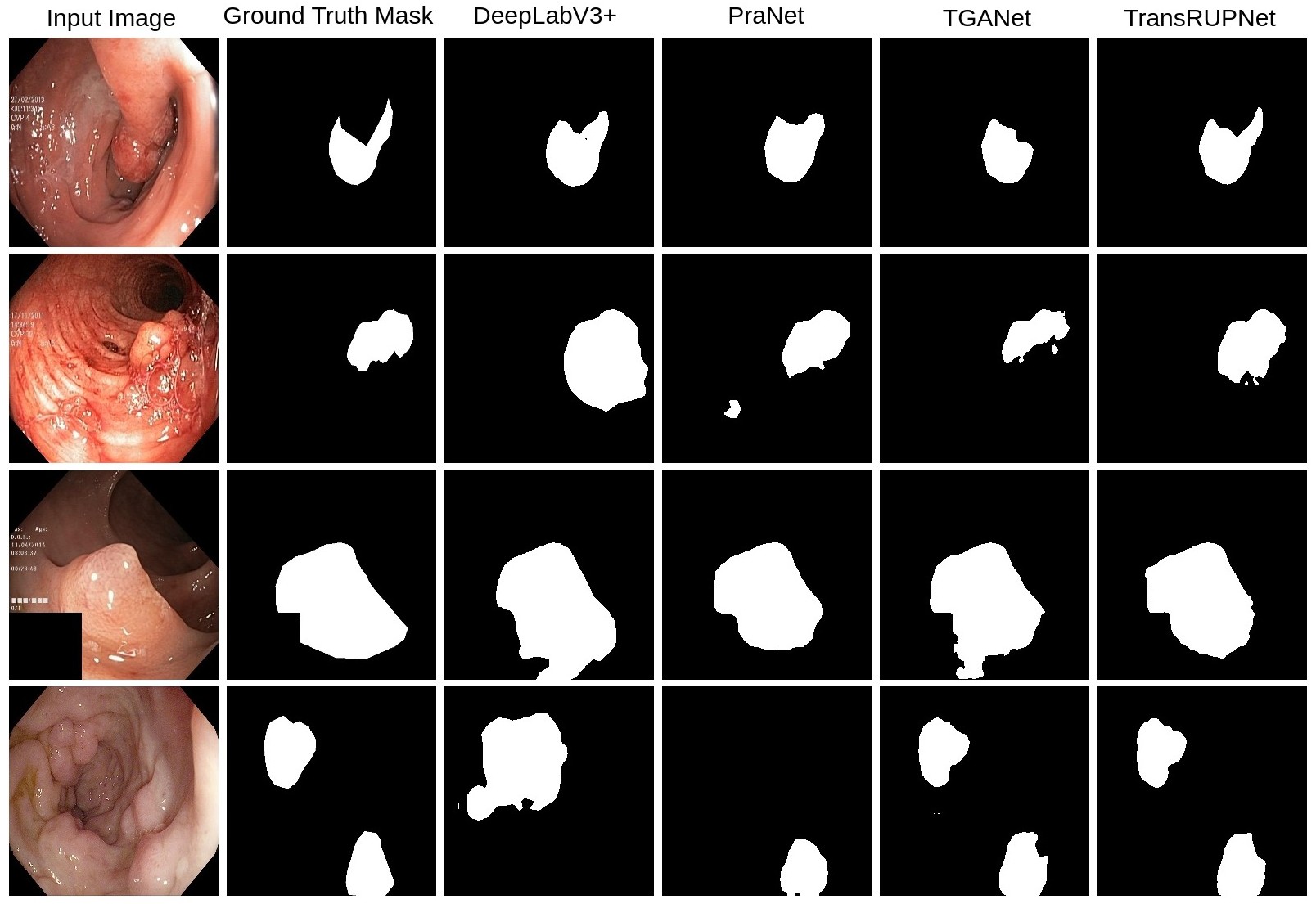} 
\caption{Qualitative example showing polyp segmentation}
\label{fig:qualitative}
\vspace{-6mm}
\end{figure}

Figure~\ref{fig:qualitative} shows the effectiveness of TransRUPNet in qualitative results. As evidenced by the Figure, TransRUPNet avoids issues such as over-segmentation or under-segmentation, which is observed in the case of SOTA TGANet and PRANet. Additionally, TransRUPNet accurately segments one or more polyps within the frames, even under challenging conditions. This highlights the robustness of TransRUPNet in handling complex scenarios and its ability to correctly delineate the boundaries of polyps. The performance drop of TransRUPNet compared to the in-distribution datasets is observed because there are insufficiently cleaned images in datasets, such as PolypGen (C6), that show elongated black regions on the left side, leading to distorted resizing and decreased OOD performance. Additionally, there are huge variations between the training dataset and OOD datasets. For instance, BKAI-IGH also contains images from FICE  (Flexible spectral  Imaging  Color  Enhancement),  BLI  (Blue Light Imaging), and LCI (Linked Color Imaging), in addition to WLI (White Light Imaging), which are not present in the training datasets. In the case of CVC-ClinicDB, it is a video sequence dataset, whereas our model is trained on still frames, which might have affected the performance. However, the performance for all the datasets is satisfactory, considering the OOD nature of the experiment.   
\vspace{-1mm}
\section{Conclusion}
In this study, we proposed TransRUPNet architecture by leveraging a pre-trained Pyramid Vision Transformer (PVT) as an encoder and incorporating a simple residual block for accurate polyp segmentation. The experimental results on various in-distribution and OOD datasets demonstrate that TransRUPNet can provide real-time feedback with high accuracy and perform significantly well on OOD datasets compared to the existing methods. By addressing the challenge of OOD generalization and providing reliable polyp segmentation results, TransRUPNet can be the strong benchmark for developing computer-aided diagnostic support systems in colonoscopy. In the future, we plan to collect more datasets from different parts of the world and build a foundational model for polyp segmentation and detection in colonoscopy.

\subsection*{Conflicts of Interest}
 The authors have no relevant financial or non-financial interests to disclose.

\section*{Acknowledgments}
The project is supported by the NIH funding: R01-CA246704, R01-CA240639, U01-DK127384-02S1, and U01-CA268808. 

\bibliographystyle{unsrt}  
\bibliography{references}

\end{document}